\documentclass[twocolumn,showpacs,showkeys,prd,floatfix]{revtex4}
\usepackage{ifthen,graphicx}
\usepackage{bm}
\usepackage{amsmath,amssymb}

\newcommand{\dotb}[1]{\hspace{-6pt} \mbox{
\rule{0pt}{12pt}  \raisebox{3.0mm}[0pt][0pt]
{ .\hspace{-0.8pt}.} \hspace{-14pt} $#1$}}

\newcommand{\dotc}[1]{\hspace{-6pt} \mbox{
\rule{0pt}{12pt}  \raisebox{3.0mm}[0pt][0pt]
{ .\hspace{-0.8pt}.\hspace{-0.8pt}.} \hspace{-14pt} $#1$}}

\begin{document}
\title{Attraction and Repulsion in Conformal Gravity}

\date{\today}

\author{Peter~R.~Phillips}
\affiliation{Department of Physics, Washington University, St.~Louis,
MO 63130 }
\email{prp@wuphys.wustl.edu}

\begin{abstract}
We use numerical integration to solve the field equations of conformal
gravity, assuming a metric that is static and spherically symmetric. Our
solution is an extension of that found by \citet{mann04}; it indicates, as
expected, that gravitation in this model should be attractive on small
scales and repulsive on large ones.
\end{abstract}


\maketitle



\section{INTRODUCTION}
\label{sec:intro}

Conformal gravity (CG), along with other alternate theories of gravity, has
been described in a review by \citet{mann6}; this paper will
referred to as PM. There have been important developments since this review
was written \citep{mann2012}, but they will not concern us here.

CG is based on the conjecture that at the deepest level the laws of nature
should be conformally invariant. Gravitation theory is therefore based on an
action principle derived from the Weyl tensor, $W^{\mu \nu}$ (PM, sections
5.1 and 8.7). An early achievement in the theory of CG was the discovery by
Mannheim and Kazanas of the analytic solution of the field equations
\begin{equation}
W^{\mu \nu} = 0
\label{eq:mannfield}
\end{equation}
for a static, spherically symmetric source, analogous to the Schwarzschild
solution in conventional gravitation (\citet{mann04}; this paper will be
referred to as MK. See also PM, section 9.2). The line element has the form
\begin{equation}
ds^2 = -B(r) dt^2 + \frac{dr^2 }{B(r)} + r^2 d \Omega_2 
\label{eq:linem}
\end{equation}
in the notation of \citet{wein2}.

MK showed that outside a static, spherically symmetric
source the function $B(r)$ is given by
\begin{equation}
B(r) = 1 - \frac{\beta(2 - 3 \beta \gamma )}{r} - 3 \beta \gamma
+ \gamma r - k r^2
\label{eq:mannsol}
\end{equation}
The final $kr^2 $ term is important only at cosmological distances, and
serves to embed the metric in a de Sitter universe. The term proportional to
$1/r$ is analogous the the Schwarzschild solution. The $\gamma r$ term is
new, and has been used by Mannheim and collaborators \citep{manobr} to
explain galactic rotation curves without recourse to dark matter.

Mannheim has also constructed a conformally invariant cosmological model
(PM, sections 3.5, 8.7, 10). In addition to the usual fermion fields, he
introduces a scalar field, $S$, that undergoes a symmetry breaking
transition and acquires a constant non-zero vacuum expectation value,
$S_0 $. The resulting field equations are (PM, equation (188)):
\begin{equation}
4 \alpha_g W^{\mu \nu} = T^{\mu \nu}
\label{eq:mannfield2}
\end{equation}
where $\alpha_g$ is a dimensionless coupling constant, and the
energy-momentum tensor, $T^{\mu \nu}$, is given in PM, equation (66):
\begin{eqnarray}
T^{\mu \nu}  & = & i \overline{\psi} \gamma^{\mu} (x) \left[
\partial^{\nu} + \Gamma^{\nu} (x) \right] \psi
- \frac{1}{4} g^{\mu \nu} h S_0 \overline{\psi} \psi \nonumber \\
  & & {} - \frac{1}{6} S_{0}^{2} \left( R^{\mu \nu}
- \frac{1}{4} g^{\mu \nu} R^{\alpha}_{\,\,\alpha} \right)
\label{eq:enmom}
\end{eqnarray}
The fermion contribution will not concern us here, because we will deal only
with the vacuum equations. But we note that even if the fermion fields are
zero, $T^{\mu \nu}$ is not zero, but depends on $S_0 $ and $R^{\mu \nu}$:
\begin{equation}
T^{\mu \nu} = - \frac{1}{6} S_{0}^{2} \left( R^{\mu \nu}
- \frac{1}{4} g^{\mu \nu} R^{\alpha}_{\,\,\alpha} \right)
\label{eq:enmomvac}
\end{equation}

In developing this cosmological model, Mannheim points out that for the FRW
space normally assumed, $W^{\mu \nu}$ is identically zero, so the field
equations are simply
\begin{equation}
T^{\mu \nu} = 0\;,
\label{eq:mannfield3}
\end{equation}
identical to the equations of conventional cosmology but with an effective
gravitational constant that is negative (PM, equation (224)):
\begin{equation}
G_{\rm eff} = -\frac{3 c^3 }{4 \pi S_{0}^{2}}
\label{eq:effg}
\end{equation}

\section{Structure of the field equations for a static, spherically
symmetric source}

We start with the observation that the solution obtained by 
MK for a static, spherically symmetric source did not use the field
equation (\ref{eq:mannfield2}) but the simpler (\ref{eq:mannfield}).

Combining (\ref{eq:mannfield2}) with (\ref{eq:enmom}), the complete field
equations can be written
\begin{equation}
W^{\mu \nu} + \eta \left( R^{\mu \nu}
- \frac{1}{4} g^{\mu \nu} R^{\alpha}_{\,\,\alpha} \right) = 0
\label{eq:compfield}
\end{equation}
where $\eta = S_{0}^{2}/(24 \alpha_{g})$ is a constant of dimension
${\rm length}^{-2}$. We shall assume the magnitude and sign of $\eta$ are
completely unknown, and for the purposes of this paper can be chosen as we
wish. We note, however, that some numerical work on the matching of interior
and exterior solutions \citep{woodmor} suggests that $\eta$ is positive.
We will call the first term in (\ref{eq:compfield}) the ``main term'', and
the rest the ``eta terms''. Since $\eta$ is not dimensionless, we can expect
that the main term will dominate for small $r$, and the eta terms for large
$r$.

If the eta terms alone are set to zero we get the equation
\begin{equation}
R^{\mu \nu} - \frac{1}{4} g^{\mu \nu} R^{\alpha}_{\,\,\alpha} = 0
\label{eq:etaonly}
\end{equation}
which leads to the same line element as before, (\ref{eq:linem}), with a
metric function of the form $B(r) = 1 + a/r + br^2 $, without the linear
term of MK.

This leads to the conjecture that at small scales the solution of MK is
appropriate, but that at some distance of order $\eta^{-1/2}$ the solution
goes over to one of Schwarzschild form. The main purpose of this paper is to
demonstrate, by numerical integration, that a solution of that kind does,
indeed, exist.

\begin{table}
\begin{center}
 
\begin{tabular}{|l|c|l|} \hline
Parameter	& Value 	& Units \\ [2pt] \hline
$r_{\rm surface}$	& 1.0E20	& ${\rm m}$ \\ [2pt]
$r_{\rm max}$	& 1.0E26	& ${\rm m}$ \\ [2pt] 
$r_{\rm center}$	& 1.0E23	& ${\rm m}$ \\ [2pt] 
$\eta$	& -1.0E-46	& ${\rm m}^{-2}$ \\ [2pt]
$\beta$	& 5.39E13	& ${\rm m}$ \\ [2pt]
$\gamma$	& 1.97E-28	& ${\rm m}^{-1}$ \\ [2pt]
$k$	& (-1.42E-51)	& ${\rm m}^{-2}$ \\ [2pt] \hline
\end{tabular}

\caption{\label{tab:params2} 
Parameters used in the integration. The second column gives values
appropriate for the galactic cluster NGC 3198. $r_{\rm surface}$ is the
disc radius, and $r_{\rm max}$ is chosen to be a typical cosmological
distance. $r_{\rm center}$ is the geometric mean of $r_{\rm surface}$ and
$r_{\rm max}$. $\eta$ is set equal to $-1/r_{\rm center}^{2}$. The value
for $k$ is enclosed in parentheses because it is not specified at the
beginning, but is determined during the fitting process.}

\end{center}
\end{table}

\begin{figure}
\centering
\includegraphics[scale=0.32]{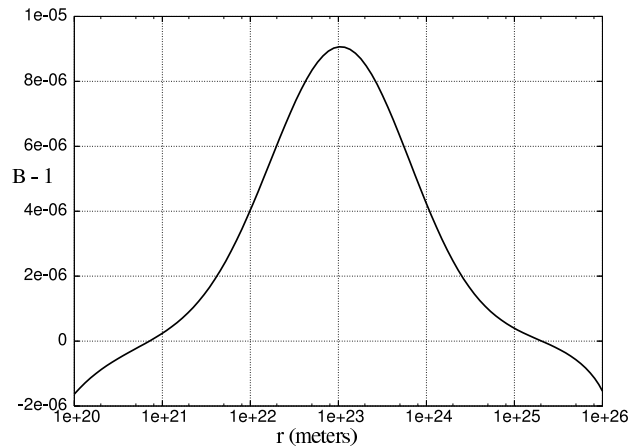}
\caption{\label{fig:j17mod} $B-1$ vs $r$, using outward integration with
$\log(r/r_{\rm center} )$ as the independent variable. This graph uses
parameters for NGC3198, as listed in Table~\ref{tab:params2}.}
\end{figure}

\section{The integration: assumptions and parameters}

As shown in table \ref{tab:params2}, we use parameters approximately equal
to those given by Mannheim for NGC3198 \citep{manobr}. Integration is
carried out over six orders of magnitude, from $r_{\rm surface}$ to a
cosmological distance, $r_{\rm max}$. 

The line element has the form (\ref{eq:linem}) at the beginning, when the
main terms of the field equation dominate, and at the end, when the eta
terms dominate. In the transition region we cannot be so sure, and a more
exact treatment might use the more general line element with two unknown
functions:
\begin{equation}
ds^2 = -B(r) dt^2 + C(r) dr^2 + r^2 d \Omega_2 
\label{eq:linem2}
\end{equation}
For simplicity, however, we will assume that, for weak fields, the form
(\ref{eq:linem}) is adequate throughout.

The tensor equation (\ref{eq:mannfield2}) gives three different equations,
but because of the relations between them it is only necessary to integrate
one. The $rr$ component yields the extension of MK (14); primes denote
differentiation with respect to $r$:
\begin{eqnarray}
B^{-1} \left[ W^{rr} + \eta \left(R^{rr} 
- \frac{1}{4} g^{rr} R^{\alpha}_{\,\,\alpha} \right)
\right] & = & \hspace{1in} \nonumber \\
  & & \hspace{-2.0in} \frac{1}{6} B^{\prime} B^{\prime \prime \prime} 
-\frac{1}{12} \left(B^{\prime \prime}\right)^2 
- \frac{1}{3r} \left(B B^{\prime \prime \prime}
- B^{\prime} B^{\prime \prime} \right) \nonumber \\
  & & \hspace{-2.0in} {} - \frac{1}{3r^2} \left[B B^{\prime \prime}
+ \left( B^{\prime} \right)^2 \right] + \frac{2}{3r^3} B B^{\prime}
- \frac{B^2}{3 r^4 } + \frac{1}{3r^4 } \nonumber \\
  & & \hspace{-2.0in} {} + \frac{\eta}{4} \left( B^{\prime \prime}
+ \frac{2}{r^2 } - \frac{2 B}{r^2 } \right) 
\label{eq:mann14}
\end{eqnarray}
Setting the right side equal to zero we get a third order equation that can
be integrated straightforwardly.

The linear term in the solution of MK is sufficiently small in practice that
$B(r)$ never deviates far from unity. We therefore integrate the equation
for the function $A(r) = B(r) - 1$:
\begin{eqnarray}
A^{\prime \prime \prime} & = & \left\{
- \frac{1}{12} \left(A^{\prime \prime} \right)^2 
+ \frac{1}{3r} \left( A^{\prime} A^{\prime \prime} \right) \right.
\hspace{1.2in} \nonumber \\
  & & \hspace{-0.5in} \left. {} - \frac{1}{3r^2} \left[
\left( A + 1 \right) A^{\prime \prime}
+ \left( A^{\prime} \right)^2 \right]
+ \frac{2}{3r^3} \left( A + 1 \right) A^{\prime} \right. \nonumber \\
  & & \hspace{-0.5in} \left. {} - \frac{A (A + 2)}{3r^4}
+ \frac{\eta}{4} \left( A^{\prime \prime}
- \frac{2A}{r^2} \right) \right\} /
\left( \frac{A + 1}{3r} - \frac{A^{\prime }}{6} \right)
\label{eq:Aref}
\end{eqnarray}

Some details of the integration are given in the appendix. Here we simply
present figure \ref{fig:j17mod}. In the fitting process, the parameter $k$
has been varied to make the curve reach $-1.0{\rm E-}6$ at $r_{\rm max}$.
We have arranged for this small $r^2$ component at the end to match a
de Sitter space; at the right of the graph, the curve is headed for a
horizon at $r = 1.0{\rm E}29$, where $A=-1\;{\rm and}\;B=0$. We note that
$|k| \ll |\eta|$, a necessary condition if the cosmological term is only
to become noticeable for $r > r_{\rm center}$.

We see, first, that for suitable choices of $\eta$ and $k$, a solution of
the field equations does exist that duplicates the solution of MK for small
$r$, and gives a Schwarzschild-like solution for large $r$. Further, we find
that $\eta$ should be chosen negative. If we change the sign of $\eta$,
we find that even for values of $|k|$ as large as $1/r_{\rm center}^{2}$ the
final value of $A$ is always large and negative.

\section{Comments}

Our graph indicates an attractive force for $r < r_{\rm center}$, and a
repulsive force for $r > r_{\rm center}$, in accordance with the well-known
feature of Mannheim's model, that gravitation is apparently attractive on
small scales but repulsive on large ones.

Checking figure \ref{fig:j17mod} against observations may prove difficult,
since so much of the figure corresponds to distances greater than
$1.0{\rm E}22\, {\rm m}$. We emphasize that in this paper, for simplicity, we
have assumed space to be free of matter for $r > r_{\rm surface}$.
\citet{manobr} have pointed out, however, that distant matter can have an
observable effect on galactic rotation curves, and our analysis would have
to be extended to take this into account. 

As a first estimate of how distant matter might affect galacic rotation
curves if figure \ref{fig:j17mod} is appropriate, let us consider a
simplified model in Minkowski space. Suppose sources within a distance of
$r_{\rm center}$ produce a linear gravitational potential, $kr$, with $k>0$.
Summing over all these sources, and approximating the sum by an integral,
we can show straightforwardly that they produce a potential near the origin
that goes like $c_1 r^2$, wth $c_1 > 0$. In other words, a test particle
released a short distance from the origin will be drawn back towards it.

Suppose in addition that sources more distant than $r_{\rm center}$ 
generate a potential more suitable for the right half of figure
\ref{fig:j17mod}, namely a Newtonian potential, $M/r$, added to a
cosmological component, $\kappa r^2$, with $\kappa < 0$. Imagine that
these sources are distributed uniformly out to some horizon, $r_{\rm max}$. 
The integrated effect of these sources near the origin can be shown to be a
potential $c_2 r^2$, with $c_2 < 0$.

We see that local and distant sources have opposite effects, and determining
which will prevail is a delicate matter, requiring a full relativistic
treatment, This will not be attempted here. We observe, however, that if,
after the $S \rightarrow S_0 $ transition, there is a time interval during
which the horizon distance is smaller than $r_{\rm center}$, then all
sources will be attractive, and we may regain the type of cosmology
familiar from Einsteinian relativity.

\vfill



\bibliography{msep}

\appendix

\section{Details of the integration}

We use the integrator RADAU5, as described in \cite{hairer}

For initial conditions we use (\ref{eq:mannsol}), the solution of MK,
evaluated at $r_{\rm surface}$.

Rather than use equation (\ref{eq:Aref}) directly, we transform it in two
ways. First, we change the independent variable to
$z \equiv \log(r/r_{\rm center})$. Using dots to denote differentiations
with respect to $z$, the equation becomes
\begin{eqnarray}
\dotc{A} & = & -2\dot{A} + 3 \dotb{A}
+ \left[\frac{1}{4(A+1) - 2 \dot{A}} \right] \times \nonumber \\
  & & {}  \left\{ - \left( -\dot{A} + \dotb{A} \right)^2 
+ 4\dot{A} \left( -\dot{A} + \dotb{A} \right) \right. \nonumber \\
  & & \left. {} - 4\left[ (A+1) \left(-\dot{A} + \dotb{A} \right)
+ \dot{A}^2 \right] \right.  \nonumber \\
  & & \left. {} + 8 (A+1)
\dot{A} - 4 A(A+2) \right. \nonumber \\
  & & \left. {} + 3 \eta r_{0}^{2} \exp(2z) \left(
-\dot{A} + \dotb{A} - 2A \right) \rule{0pt}{16pt} \right\}
\label{eq:Afullz}
\end{eqnarray}

The second transformation is simply a scale change. $A(r)$ is always much
less than unity, and its derivatives are smaller still. We can get more
manageable magnitudes by multiplying all variables by a constant, $P$,
chosen so that $PA(r_{\rm surface}) = 1$. The scale factor is removed after
the integration is complete. Writing $A_P = PA$, $\dot{A}_P = P \dot{A}$,
etc.,
\begin{eqnarray}
\dotc{A_P } & = & -2\dot{A}_P + 3 \dotb{A_P }
+ \left[\frac{1}{4(A_P +P) - 2 \dot{A}_P } \right] \times
\hspace{0.5in} \nonumber \\
  & & {}  \left\{ - \left( -\dot{A}_P + \dotb{A_P } \right)^2 
+ 4\dot{A}_P \left( -\dot{A}_P + \dotb{A_P } \right) \right. \nonumber \\
  & & \left. {} - 4\left[ (A_P +P) \left(-\dot{A}_P + \dotb{A_P } \right)
+ \dot{A}_{P}^2 \right] \right.  \nonumber \\
  & & \left. {} + 8 (A_P +P)
\dot{A}_P - 4 A_P (A_P +2P) \right. \nonumber \\
  & & \left. {} + 3P \eta r_{0}^{2} \exp(2z) \left(
-\dot{A}_P + \dotb{A_P } - 2A_P \right) \rule{0pt}{16pt} \right\}
\label{eq:AfullzP}
\end{eqnarray}

(Because $A(r)$ is always so small, we can use either this complete
equation or its linearized form; the resulting graphs are very similar.)

Before beginning the main integration, we find two values of $k$, one of
which drives $A(r)$ to a large positive value at $r_{\rm max}$, while the
other drives $A(r)$ to a large negative value. These two values are used as
input to a root-finding program, such as {\tt rtbis} from Numerical Recipes
\citep{numrec}. Each pass through the root finder involves a complete
integration, with the output being the value of $A(r_{\rm max} )$. $k$ is
automatically adjusted in this way to bring $A(r_{\rm max} )$ to the chosen
final value.
 
\end{document}